# The management, treatment and study of skeletal pain harnessing tissue engineering models.

Lucia Iafrate[1], Maria Cristina Benedetti[1,2], Samantha Donsante[3], Alessandro Rosa[1,2], Alessandro Corsi[3], Richard OC Oreffo[4], Mara Riminucci[3], Giancarlo Ruocco[1], Chiara Scognamiglio[1], Gianluca Cidonio[1,4]*

[1] Center for Life Nano- & Neuro-Science (CLN$^2$S), Fondazione Istituto Italiano di Tecnologia, Rome, Italy

[2] Department of Biology and Biotechnologies "Charles Darwin", Sapienza University of Rome, Piazzale Aldo Moro 5, 00185 Rome, Italy

[3] Department of Molecular Medicine, Sapienza University of Rome, Italy

[4] Bone and Joint Research Group, Centre for Human Development, Stem Cells and Regeneration, Institute of Developmental Sciences, University of Southampton, Southampton, United Kingdom

*Corresponding author. Gianluca Cidonio, PhD.

Tel.: +39 06 49255207

*Email address*: gianluca.cidonio@iit.it,

# Abstract

Bone pain typically occurs immediately following skeletal damage with mechanical distortion or rupture of nociceptive fibres. The pain mechanism is also associated with chronic pain conditions where the healing process is impaired. Any load impacting on the area of the fractured bone will stimulate the nociceptive response, necessitating rapid clinical intervention to relieve pain associated with the bone damage and appropriate mitigation of any processes involved with the loss of bone mass, muscle, mobility and, to prevent death.

The following review has examined the mechanisms of pain associated with trauma or cancer-related skeletal damage focusing on new approaches for the development of innovative therapeutic interventions. In particular, the review highlights tissue engineering approaches that offer considerable promise in the application of functional biomimetic fabrication of bone and nerve tissues. The strategic combination of bone and nerve tissue engineered models provides significant potential to develop a new class of *in vitro* platforms, capable of replacing *in vivo* models and testing the safety and efficacy of novel drug treatments aimed at the resolution of bone-associated pain. To date, the field of bone pain research has centred on animal models, with a paucity of data correlating to the human physiological response. This review explores the evident gap in pain drug development research and suggests a step change in approach to harness tissue engineering technologies to recapitulate the complex pathophysiological environment of the damaged bone tissue enabling evaluation of the associated pain-mimicking mechanism with significant therapeutic potential therein for improved patient quality of life.

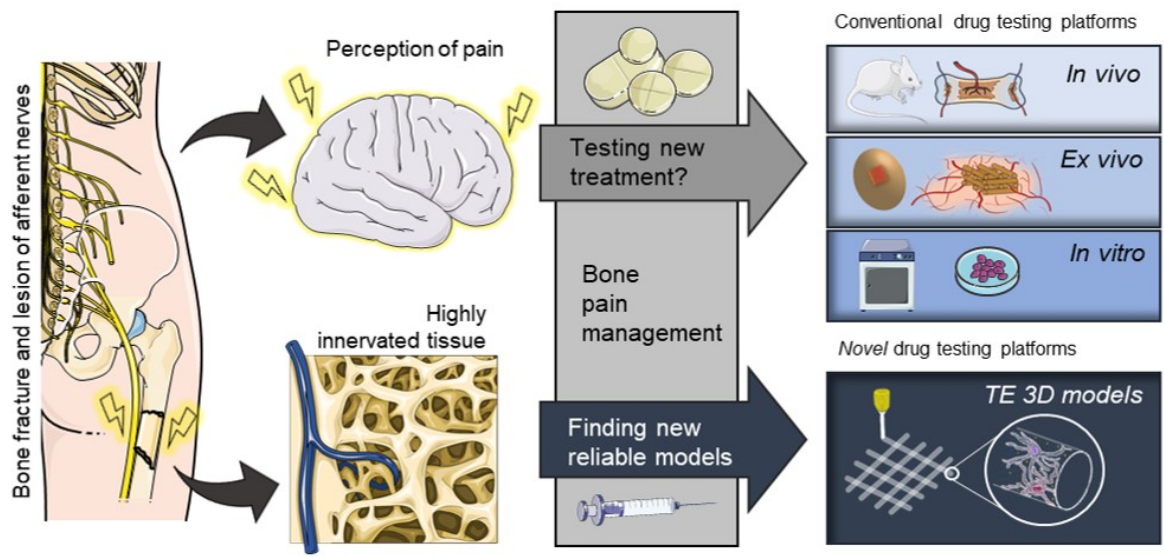

**Graphical Abstract.** Rationale underlying novel drug testing platform development. Pain detected by the central nervous system and following bone fracture cannot be treated or exclusively alleviated using standardised methods. The pain mechanism and specificity/efficacy of pain reduction drugs remain poorly understood. *In vivo* and *ex vivo* models are not yet able to recapitulate the various pain events associated with skeletal damage. *In vitro* models are currently limited by their inability to fully mimic the complex physiological mechanisms at play between nervous and skeletal tissue and any disruption in pathological states. Robust innovative tissue engineering models are needed to better understand pain events and to investigate therapeutic regimes

# 1. Introduction

The increase in life expectancy as a consequence of medical advances over the past decades has also heralded an emergence of pathologies correlated to the ageing demographic [1, 2]. Acquired skeletal diseases such as *osteoporosis*, *cancer metastasis* and *multiple myeloma,* and genetic disorders such as *osteogenesis imperfecta* and *fibrous dysplasia* of bone are known to cause bone fragility as a consequence of osteopenia or locally enhanced bone resorption [3, 4]. Thus, skeletal disorders often lead to bone fractures, resulting in extremely painful consequences for the patient both during movement or at rest [5]. For instance, the initial skeletal damage elicited by primary bone tumours or metastasis typically facilitate the fracture of diseased bone tissue with a consequent inflammation response and sudden pain. During the progress of the pathological state, cancer bone pain becomes more perceptible becoming progressively constant and intense [5]. Pathological pain is generally perceived at the site of injury or often as referred pain with muscle spasms together with an extensive range of action, involving multiple sites other than the site of lesion [6, 7].

As well as pathological conditions, bone fracture/trauma are often a consequence of damage as a result of sporting activities [1]. Sports-related fractures, especially in the upper limbs are currently the third leading cause of bone fractures, affecting a wide demographic including the young [1, 8]. Furthermore, reduced physical activity, poor nutrition (such as limited Vitamin $D_3$ intake) and sedentary lifestyles can impact on bone structure, increasing the risk of bone fracture and impaired skeletal tissue composition [9–12]. There remains an unmet need for improved pain management programmes to: (i) ease the pain, (ii) improve tissue healing and, (iii) increase the quality of the life of hospitalized patients [13]. To date, the complexity of the pain process has limited clinical options, as current therapies often aim to alleviate the discomfort rather than remove or resolve the pain [13].

The current review explores the pathophysiology and the processes involved in pain mechanisms affecting bone tissue, from central causes related to pathological and damage-related pain to the evaluation of current approaches and therapeutic analysis. Furthermore, the review examines currently available functional *in vivo* and *in vitro* models, highlighting the most effective platforms involved in delineating the pain processes. Ultimately, a focus on current research in the skeletal pain modelling platforms is presented and the rich vistas of opportunity therein together with a detailed evaluation of future trends and perspectives for the engineering of *in vitro* skeletal disruption and pain.

## 2. Pathophysiology of cancer and trauma fracture bone pain

Unlike other types of physical discomfort, the pathogenesis of bone pain is still poorly understood due to the complexity of the underlying functional mechanisms [6]. Indeed, pain affecting musculoskeletal system does not occur simply through a mechanical distortion of the nociceptive fibres but involves a plethora of supplementary mechanisms [13] (**Fig. 1**).

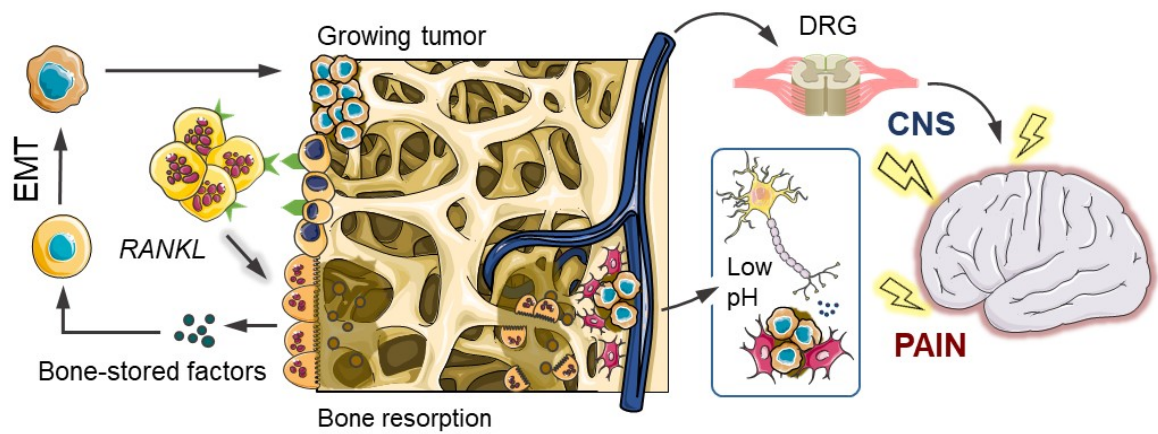

**Fig. 1** Bone pain mechanism. Bone pain cycle between osteoclasts, cancer cells and bone fractures in osteolytic metastases. Bone-derived growth factors promote proliferation and stimulate epithelial–mesenchymal transition (EMT) of cancer cells and the production of bone-modifying cytokines, in bone-colonizing cancer cells. These factors further stimulate osteoclastic bone resorption via activation of receptor activator of the nuclear factor-kB (RANKL)/RANK pathway in osteoblasts and osteoclasts, increasing the release of bone-stored growth factors. Osteoclasts activity induces a local acidosis increasing TRPV1 and ASIC3 ion channels activity of nociceptor fibres. The response is sent to the central nervous system (CNS) via dorsal root ganglion (DRG) together with mechanical stimulation due to the growing tumour mass-induced bone fractures

## 2.1 Mechanisms of cancer bone pain

Cancer bone pain mechanisms are complex and heterogeneous typically involving elements of inflammatory and neuropathic pain, with neuro-chemical changes from peripheral, spinal and central sites [7]. The inflammatory process arising from the bone lesion produced by the growing tumour mass involves cancer cells which release pain mediators, contributing to attract additional immune cells to the lesion site further stimulating nociceptive fibres [7]. Osteolytic lesions are characterized by a rapid resorption of the bone resulting in bone fragility and subsequent fracture risk [14]. Osteoclastogenesis is promoted by cancer cells via receptor activator of nuclear factor kappa-B ligand (RANKL) [14] stimulating osteoclast maturation, with a consequent reduction of the pH at the local micro-environment [7, 15].

The local acidosis, produced by osteoclast activity and cancer cell metabolism (Warburg effect), contributes to the sensitisation of the primary afferent neurons present in the bone [16]. The high local concentration of protons activating the sensing ion channels TRPV1 and ASIC3 of sensory neurons, increasing the nociceptive response, which is subsequently translated into an acid-evoked cancer bone pain event [7, 16]. Tumour expansion results in mechanical distortion of the bone including the stretching of the periosteum, the outer envelope of the cortical bone. The periosteum holds a high surface density of mechanosensitive sensory nerve fibres which, once stimulated, transmit the nociceptive response to the brain [4, 13, 17].

Osteoblastic lesions are typically associated with pathologies such as prostate cancer metastasis, Hodgkin's lymphoma and medulloblastoma [18]. Each of these pathologies is characterised by the deposition of new bone, typically brittle in nature in comparison to the healthy bone tissue [19]. Moreover, the arrangement of collagen fibres has been reported to be random, indicative of woven bone with poor load distribution capacity and consequent risk of local (micro-) fractures and generation of bone pain [15]. To date, the mechanisms around the development of osteoblastic metastasis remain poorly understood [19]. The neuropathic component of bone pain may arise from cancer-induced damage to the sensory nerves caused by infiltration, compression by tumour cells, tumour-induced hyper-innervation and stretching or denervation as the lesion expands and the bone degrades [7, 16].

## 2.2 Mechanisms of trauma fracture bone pain

Bone fractures resulting from trauma are influenced by: i) the quality and density of the bone, ii) the mechanical property of the bone affected and iii) the application of the load [20]. During the traumatic event, especially in the case of a fall, bone is subjected to a multi-axial load condition, which can result in a fracture due to the multiple excess loads in an axial-torsion [21, 22]. Following fracture, the altered orientation of the damaged bone is perceived by mechanosensitive fibres as a mechanical stimulation [13]. The distortion of the cortical bone, or an increase in intraosseous pressure, activates the nociceptive C and A-δ fibres in the periosteum, responsible for the perception of dull pain and acute pain, respectively [23]. The initial nociceptive response to injury, as in cancer bone pain, is sent to the brain, generating the sensation of pain [1, 13, 17, 23]. Unlike cancer bone pain, the inflammatory component is due to the formation of a hematoma following the rupture of blood vessels accompanying the bone damage [24]. As in cancer bone pain, even in trauma induced fracture, inflammatory-related factors can directly activate or sensitise nociceptors [13, 25]. Following bone fracture, any movement, or applied load, determines a mechanical stimulation of the sensory nerve fibres that innervate the periosteum, the mineralized bone and the bone marrow, generating a state of neuropathic pain [13, 26].

Nerve growth factor (NGF) is involved in the sensitization and germination of nociceptor fibres [27, 28]. Following fracture, the neurotrophic factors, released by the inflammatory and stromal cells, induce an ectopic sprouting, resulting in hyper-innervation of the marrow, the mineralized bone and the periosteum. The enhanced nervous network makes any type of normal load or movement of the bone perceived as noxious [13]. Fracture realignment promotes healing with the levels of NGF and sprouted nerve fibres reduced, restoring normal bone innervation and as a consequence pain relief [13]. During chronic pain the concentration of inflammatory and neurotropic mediators increases, favouring excessive nerve sprouting leading to the formation of a neuroma-like structure, highly sensitive to any type of movement (mechanical allodynia) [1]. It is clear, fracture healing must occur as soon as possible in order that bone pain does not develop into a chronic pain condition.

## 2.3 Therapeutics and current standard treatment approaches for cancer and fracture bone pain

Optimised bone pain treatments are essential to improve patient quality of life in pathological conditions and to aid bone healing following trauma-induced bone fracture. Current pain reduction approaches include:

*Radiotherapy*: Typically applied for single metastases. The use of radiation aims to reduce the tumour mass to reduce the chemical mediators that mediate bone pain [4]. It has been found that 70% of patients experience partial pain relief within 2 weeks with only 25% of patients experiencing total pain relief [5].

*Bisphosphonates*: Bisphosphonates (BPs) are often administered in combination with radiotherapy or, in cases where the pain is not localized [5]. BPs can be used in addition to analgesics and in other metabolic bone diseases to reduce or prevent disease progression and to decrease disease symptoms and complications [3, 29, 30]. However, BP administration is accompanied by a number of side effects including flu-like symptoms (fever, arthralgia, myalgia and weakness), anaemia, nausea, dyspnoea, peripheral edema and, in rare cases, osteonecrosis of the jaw [19, 30].

*Monoclonal antibodies* (mAbs): Denosumab binds to RANKL preventing the interaction with the receptor on the osteoclasts and thus preventing osteoclast maturation and function and is one of the most successful mAbs treatment options available [5, 19, 31, 32]. Tanezumab treatment is a potential alternative to Denosumab, binding to NGF to prevent interaction with TrkA and p75 receptors [5, 33]. However, Tanezumab is still under initial clinical assessment and has not yet been cleared for safe therapeutic application [34].

*Analgesics*: Non-steroidal anti-inflammatory drugs (NSAIDs) are administered in patients with mild or moderate bone pain [3]. NSAIDs inhibit prostaglandin (PG) synthesis acting on cyclooxygenase, reducing local edema and PG-induced nociceptor sensitization and local inflammation [3, 5, 35]. Following NSAIDs therapies moderate or severe pain was reduced to mild pain after 2 weeks in 51% of patients [5]. Nevertheless, long-term treatments with NSAIDs can have significant detrimental effects on skeletal health as a consequence of inhibition of osteoblast growth due to cell cycle arrest and apoptosis induction [3, 36].

*Corticosteroids*: Dexamethasone (Dex) has been routinely used for the management of metastatic bone pain, neuropathic pain from infiltration or compression of neural structures [5, 37]. Corticosteroids exert potent anti-inflammatory effects and can directly decrease the impaired electrical activity of damaged peripheral neurons, decreasing the intensity of pain [38]. However, there remains a paucity of data on the efficacy of corticosteroids and current therapies are typically administered over limited time frames [5].

The complexity of the pain mechanism and the plethora of unsuccessful therapeutic approaches together with varied and often patient subjective response to pain, necessitate new modelling and evaluation platforms. Development of appropriate modelling platforms offer new avenues to evaluate the safety and efficacy of novel drugs that, could improve patient quality of life.

## 3. Tissue engineering models

Tissue engineering (TE) seeks to harness cells, engineering, materials together with biochemical and physico-chemical cues to restore, maintain, improve, or replace tissues of interest. Cells in combination with 3D scaffolds offer a physiologically relevant environment to examine cell fate, tissue maturation, or *in situ* regeneration as well as drug screening *in vitro, ex vivo* and *in vivo* [39, 40]. Cell approaches include incorporation of primary cells (terminally differentiated or stem cells) from patients that can be expanded *in vitro* and encapsulated in biocompatible water-based matrices, called hydrogels. Thus, biomimetic models capable of resembling the complex pathophysiological state of tissues are an important goal for the tissue engineer to generate physiologically relevant models and tissue constructs.

### 3.1 Cell sources

To closely mimic and recreate the tissue micro-environment, 3D models an appropriate cell source is pivotal. Human bone marrow stromal cells (HBMSCs) are commonly used for bone tissue engineering purposes given their ready availability. HBMSCs contain a subset of cells, skeletal stem cells, that can self-renew and differentiate into cells of the stromal lineage namely, chondrocytes, osteoblasts and adipocytes [41–43]. Differentiated skeletal populations present a window into cellular changes due to specific pathologies from the affected donor although cell numbers are naturally limited given limited expansion capacity of a differentiated cell population [42].

The limited availability and paucity of robust methodologies for the isolation of neurons human brain tissue has brought stem cells, specifically pluripotent stem cells (embryonic and induced pluripotent stem cells (ES and iPSCs)), to the fore as a pathway to derive and differentiate neuronal populations and for the generation of novel and functional *in vitro* models [43].

iPSCs have garnered significant interest given their potential as a novel cell source, avoiding crucial ethical issues, reproducibility and challenging isolation and expansion protocols. Indeed, iPSCs can be readily derived from human skin biopsies from patients and donors and with *ex vivo* induction towards an embryonic-like cell state, iPSCs can subsequently differentiate into a wide spectrum of tissues [43, 44]. iPSCs are typically differentiated into neuroectodermal and neuronal cells [44, 45], with limited proven potential to differentiate into sensory cells [44]. Terminally-differentiated cells such as dorsal root ganglion (DRG) offer an alternative to iPSCs-induced neurons for *in vitro* studies and have been used in the detection of noxious stimuli and pain mechanisms [46]. Mouse DRGs are relatively easy to dissect and culture, however, it is important to note human DRGs display different responses to analgesics compared to rodent DRGs [44]

### 3.2 Biomaterials

The mechanical properties of scaffolds at a macroscopic and microscopic scale play a crucial role in regulating cell behaviour [47] and, typically, vary dependently on the biomaterial system of choice. Hydrogels display exceptional biocompatibility, hydrophilicity, degradability and oxygen/nutrient permeability together with structural stability [48]. A number of studies have shown that mechanical properties of biomaterials influence and guide HBMSCs differentiation along the different stromal lineages depending on the particular Young's moduli (or elastic modulus, E) of the material [49]. Osteoconductivity and osteoinductivity are additional properties given a bone biomaterial platform will ideally promote bone formation and guide autologous and skeletal-residing stem cells to differentiate into bone cells [50]. In contrast, biomaterials with a low elastic modulus are required for the regeneration of neural tissues. To guide the repair of neural tissue, the biomaterial system needs to facilitate and foster neural network formation, mimicking the same function of the neural extracellular matrix (ECM) [51]. Thus, differences in Young's moduli impact on cell fate: a biomaterial with $E < 1$ kPa facilitating brain tissue development with iPSCs induced to differentiate into neurons, while E in the range of 90-230 kPa may be suitable for the spinal cord tissue development [48, 52].

### 3.3 TE 3D models and technologies

3D *in vitro* models offer novel tools for the evaluation of drug safety and efficacy. Two-dimensional culture of different cell types has allowed the generation of models able to explore diseases from a multi-tissue perspective. Nevertheless, the inability to fully recapitulate the complexity of the disease micro-environment and architectural functionality has proved detrimental and limiting. 3D models have thus in recent years attracted much interest as reliable and reproducible approaches for the fabrication of new biomimetic models for drug screening.

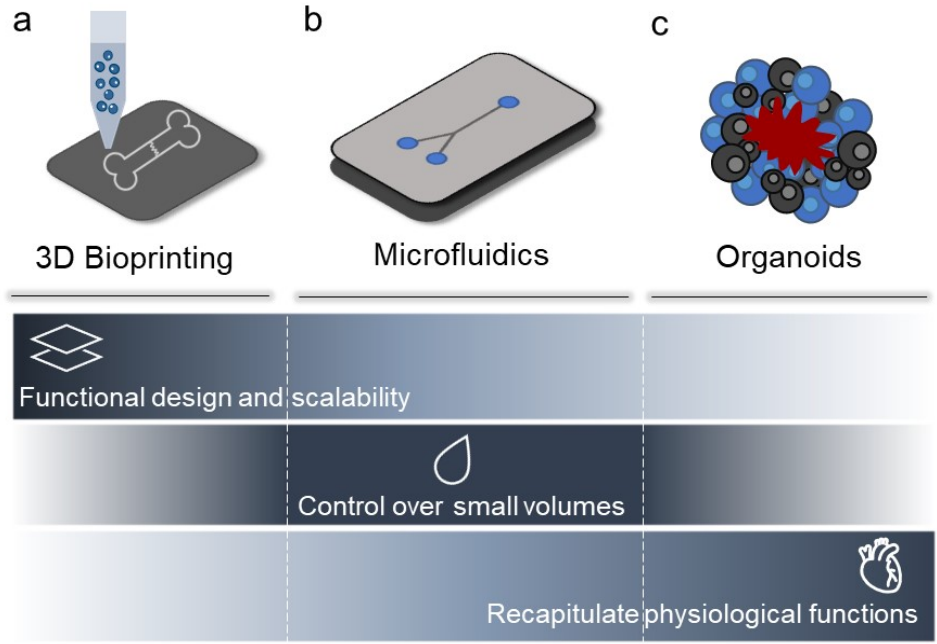

**Fig. 2** Tissue Engineering platforms for *in vitro* disease modelling. Modelling platforms for the recapitulation of bone-neuro pathological conditions include: (a) 3D Bioprinting, (b) microfluidics and (c) organoids. These systems hold great potential in mimicking the disease conditions present in bone and neural tissue. 3D Bioprinting technologies offer the ability to pattern functional architectures and design as well as the ability to 3D print scalable and complex tissues. Microfluidics, lack the above listed abilities, but can precisely control small volumes of liquid required to create compartmentalised micro-environments for the development of in vitro models. Organoids, in contrast, offer biomaterial-free approaches with application of self-assembling properties of different type of cells to build and recapitulate physiologically functional tissue substitutes/models

### 3.3.1 3D Bioprinting

Additive manufacturing (AM) technologies support the layering of materials to print objects from three-dimensional model data, layer upon layer, facilitating the manufacture of structures with a predefined geometry and size [53]. 3D bioprinting is a widely used AM technology to produce functional cell-laden scaffolds using polymeric bio-inks. [53, 54]. Generally, a 3D bioprinted scaffold for *in vitro* culture needs to present specific properties such as biocompatibility, controlled kinetics of biodegradability and comparable structural/mechanical properties to the native tissue to permit adequate oxygenation, mass transfer, nutrient exchange and vascularisation [48, 53, 54]. Thus, vascularisation, while still a major challenge, is essential to recreate the natural metabolic functions of tissues, such as nutrient transportation and waste removal [55]. Harnessing 3D bioprinting approaches, it is possible to deposit cells, layer by layer, to produce three-dimensional complex cellular structures using computer-aided design (CAD) [53, 56]. Moreover, 3D bioprinting technologies offer the advantage of building complex and hierarchical microstructures with high resolution and reproducibility, suitable for cell survival, proliferation and maturation (**Fig. 2a**) [53, 57].

### 3.3.2 Microfluidics

TE has benefited from advancements in microfluidic-based technologies, in particular from the development of organ-on-a-chip platforms [58–60]. Such devices are designed to house 3D multi-cell culture within interconnected channels and/or chambers, where the transport of nutrients and biochemical cues is precisely controlled by laminar flows. Specifically, the geometry of the device combined with the tuning of fluidic parameters allows for fine spatiotemporal patterning of chemical, physical, and mechanical properties, resulting in the reproduction of a biomimetic tissue microenvironment. The ability to integrate micro-actuators, and the application of external stimuli (e.g., mechanical [61], acoustic [62] and electrical [63]), have enhanced the ability to recreate pathophysiological conditions, providing microfluidic devices as a superior alternative to conventional *in vitro* models.

High-throughput analysis, enhanced reproducibility as well as cost reduction are additional advantages of microfluidic platforms [64]. However, the use of microfluidic devices for cell culture is still not widespread due the challenging fabrication procedures and the expensive instrumentation typically required. Moreover, microfluidic platforms suffer from poor scalability and can only partially replicate, at the micro-scale, the pathophysiological damage at the tissue macro-scale level relevant for clinical applications (**Fig. 2b**) [48].

*3.3.3 Organoids*

Organoids are three-dimensional *in vitro* cell models that consist of clusters of cells, such as iPSCs, ESCs or adult stem cells (ASC) that self-organise spatially and differentiate into functional mini-organs using a scaffold-free approach. The mature structures are able to recapitulate, to some degree, the cellular composition, architecture and functions of a native organ [48, 65–67], thus used for *in vitro* disease modelling (**Fig. 2c**). Organoids are particularly useful for the identification and testing of new therapeutic treatments given organoids can adapt to any *in vitro* genome modification or gene therapy [66]. To date, reproducibility has proved a major issue, due to significant intra- and inter-batch differences in terms of size and cell organization and composition. Furthermore, organoids lack vascular perfusion, mechanical signals, long-term stability and circulating immune cells, essential for physiological function [66, 68].

# 4. Towards development of bone pain models

Currently, animal models are the most widely used platform for drug discovery and screening, despite significant issues associated with reproducibility, availability and ethical considerations [9]. An alternative is provided by *ex vivo* models: explants of human or animal tissue that contain the cellular and extra-cellular composition needed to replicate the *in vivo* conditions [9] and can be applied for the screening of new drugs for safety and efficacy [69]. The following section details the most effective models for simulating bone and neuro tissues, as well as their close interaction within a pathological state.

## 4.1 Bone models

Bone is a complex dynamic tissue, capable of regeneration following damage. However, the majority of pathologies associated with skeletal tissue remain unresolved with TE addressing some of the challenges through development of viable models for the recapitulation and study of such skeletal pathologies.

*4.1.1 In vivo*

*In vivo* bone models allow the examination of pathologies with a comparable degree of cell, biophysical and biochemical signalling observed in the human body [70, 71]. The selection of a specific *in vivo* skeletal model is not without challenges from consideration of the pathophysiology of the skeletal disease, species differences, timelines and variations in bone properties (density, hardness, architecture, porosity and bone turnover) across species [72] [70]. *Small* animal models (rodent) for bone-related diseases (**Fig. 3a**) are widely used given the associated low cost and their accelerated metabolism [73]. Rodents and small animals are generally indicated for the study of bone metabolism and regeneration related to age, fractures, osteoporosis and osteoarthritis [74–76]. *Large* animal pre-clinical models, such as sheep [77], dogs [78] and pigs [79], are often used to study size-related processes or metabolic characteristics, which can be comparable to the human pathophysiology, as in the case of long-term diseases, biomechanics or bone healing efficiency. The domestic sheep has found application as a model for orthopaedics and traumatology research given their similar weight to the human body [70]. In terms of organic, inorganic volatile fraction, water and ash content, studies indicate canine models are able to recapitulate the human physiological state [71, 80], and thus have found application in examination of joint disorders, especially osteoarthritis [70]. Nevertheless, the use of preclinical animal models is currently widely accepted as replaceable, with extensive effort in providing new alternatives to reduce and ultimately, replace animal models.

### 4.1.2 In vitro

*In vitro* models using stem cells can generate specific tissue platforms with the appropriate conditioning and *in vitro* stimulation with HBMSCs has been widely used as a building block for 3D complex constructs that aim to explore bone repair and regeneration.

### 4.1.3.1 3D Bioprinting

The use 3D bioprinting of bioactive and biodegradable materials in combination with stem cells has garnered significant interest for the fabrication of functional skeletal models. Natural and synthetic biomaterials have been extensively explored for the printing and modelling of diseased bone tissue. However, as detailed above, these approaches still cannot recapitulate the complex skeletal micro-environment [81, 82]. An alternative is provided by the use of a combination of polymeric biomaterials with calcium phosphate cements [83], hydroxyapatite [84], β tri-calcium phosphate (β-TCP) [85] or bio-glasses [86], to produce composite scaffolds improving the binding interaction and mechanical properties of the material during the 3D printing process. Such composites offer, potentially, an initial mineralized osteoinductive, as well as osteoconductive, support for encapsulated skeletal cells [30, 54, 82, 87]. Recent studies have examined the use of bioinks with nanocomposites such as the synthetic nanosilicate clay (Laponite® (LAP)) [88]. LAP displays a capacity to promote the differentiation of HBMSCs in osteogenic cells both *in vitro* (**Fig. 3b**) and *in vivo* [89] and can be used in combination with polymeric materials to promote cell viability. Cidonio and co-workers [90] demonstrated the combination of nanoclay with a library of materials including gellan gum (GG) [90], gelatin methacryloyl (GelMA) [91] and alginate-methycellulose [88, 92] to produce functional bioinks to model bone tissue. While a range of material approaches exists for bone scaffold manufacture, to date, the recapitulation of the natural bone micro-environment has proved elusive.

### 4.1.3.2 Microfluidic

Osteoclasts and osteocytes have been established in co-culture in microfluidic devices in the presence of fluid flow stimulation to explore the mechanotransduction and intercellular signalling pathways involved in osteoporosis [93]. Recently, Ma and colleagues have developed a model of rheumatoid arthritis (RAs) in a microfluidic chip to investigate fibroblast-like synoviocyte-mediated bone erosion. When cultured in the RA-mimicking chip, human synovium cells were found to exhibit higher migration potential, associated with enhanced expression of cadherin-11, towards the co-culture of RANKL-stimulated osteoclastic RAW264.7 and BMSCs, confirming the modelling ability of the microfluidic platform [94]. The microfluidic approach has been found particularly suitable for the study of bone tumour invasion [95], metastatic intravasation [96] and extravasation [97]. Conceição and co-workers presented a humanized organ-on-a-chip model of the breast cancer bone metastatic niche [98]. The microfluidic device housed three interconnected chambers including neuro, breast and bone cells, respectively. The synergistic interplay between neurons and osteoclasts towards bone tropic breast cancer resulted in increased levels of pro-inflammatory cytokines cells.

Despite the great versatility of the microfluidic approach, a crucial limitation remains the poor level of biomaterial mimicry. Thus, the absence of rigid matrices in microfluidic chips limiting the reproduction of the complex calcified ECM found *in vivo* are limiting bone TE research.

4.1.3.3 *Organoids*

Bone-cartilage organoids (BCORG), representing "mini-joint" models, have been obtained by co-culturing cells isolated from paediatric rib tissues containing both bone and cartilage in osteochondral medium [99]. Despite limitations, in terms of ultimate maturation, BCORG may provide a tool for osteoarthritis disease modelling and drug testing. Abraham and colleagues, reported that joint organoids were partially responsive to treatment with adenosine $A_{2A}$ receptor agonists, previously used in murine models of osteoarthritis to reduce skeletal tissue damage [99]. Organoids have emerged as key players in modelling pathological states caused by an imbalance of bone tissue resorption and deposition processes, such as osteoporosis and loss of bone mass due to reduced mechanical stimulation during space exploration missions. For instance, Iordachescu and co-workers [100] cultured bone organoids, generated by seeding primary osteoblasts and osteoclasts onto femoral head micro-trabeculae, in a bioreactor simulating microgravity conditions [100]. Interestingly, compared to static controls, the simulated microgravity organoids showed altered morphology of resorption sites and reduced bone mass. Thus, organoids offer an attractive approach to functional bone tissue modelling given the ease of fabrication and relatively simple, yet long, maintenance and maturation potential. However, inconsistencies in scale up and reproduction of identical bone organoids remain a significant challenge.

## 4.3 Neuro models

The engineering of a functional nerve tissue *in vivo* or *in vitro* remains an unmet research challenge. Two- and three-dimensional cell cultures have offered valuable information on the study of nervous system diseases but remain limited in their ability to model human neural development [101]. To date, *in vivo* animal-based neuro models remain the approach of choice [102] with yet limited success in recapitulating the neural micro-environment *in vitro*.

*4.3.1 In vivo*

*In vivo* animal-based neuro models are extensively used for the study of physiological and behavioural consequences of pain-associated mechanisms [102]. Currently, animal nerve injury models offer the most promising approach to study the development of pain. Surgical nerve lesions provide information on how neuropathic pain is induced in animals, which share many similarities to human pain [102]. A seminal model was proposed by Bennett and Xie [103], using a chronic constriction injury of the sciatic nerve in rodents closely mimicking human neuropathic distress resulting from a trauma of the peripheral as well as mechanical allodynia [102]. Another *in vivo* model with lesion of the partial sciatic nerve has been developed by Seltzer and co-workers [104] inducing neuropathic pain and mechanical allodynia. Nerve injury was created by tying the third dorsal nerve at the middle of the sciatic nerve at the level of the upper thigh of the rodent [102]. Xie and colleagues [105] investigated the possibility to alleviate chronic pain experienced by inhibiting glial fibrillary acidic protein (GFAP)-positive glial cells (**Fig. 3c**). Results indicate the neuropathic pain behaviour induced by spared nerve injury (SNI) significantly decreased in mice with downregulated nuclear factor κB (NFκB). However, anatomical and physiological differences between rodent and the complex human nervous system given the marked differences in morphology, number of nerve cells, laminar distribution and gene expression limit many studies [101]. Human models would represent the optimal platform to study neuropathic pain-related pathological changes, however, limitations in the availability of post-mortem samples, biopsies, neuroimaging and neuropharmacological treatments remain [101]. Biopsies can be subject to a severe response involving reactive cells, proliferation and progressive neurodegeneration, reducing their suitability to accurately model a wide range of diseases [101, 106].

### 4.3.2 In vitro

Alternatively, *in vitro* models comprising neuronal cells are currently in development to aid the reduction, replacement and refinement (3Rs) of *in vivo* animal models for the simulation of pathophysiological pain mechanisms [44]. 3D neuro cultures have been shown to support cell differentiation, increased neuritis outgrowth and myelination [65, 107], harnessing 3D bioprinting, microfluidics and organoid systems technologies.

#### 4.3.2.1 3D Bioprinting

3D bioprinted *in vitro* neuro models, as detailed for bone tissue models, requires the choice of a suitable biomaterial able to support cell survival, proliferation, and functionality of the encapsulated cells. Functional biomaterials used for neuro tissue 3D bioprinting are typically natural polymers such as alginate [108, 109], collagen type I, silk fibroin [110], matrigel [111], chitosan [112], hyaluronic acid (HA) [111, 113], methylcellulose [114] and blends such as alginate/carboxymethyl-chitosan/agarose [112] as well as synthetic polymers poly(ethylene-glycol) (PEG) [115, 116] and functionalised materials, such as GelMA [117]. To improve cell viability, functionality and cell adhesion, these biomaterials are often functionalized with peptides such as -RGD, -IKVAV and -YIGSR, or proteins such as laminin and fibronectin [48]. Lozano and colleagues [51] proposed the printing of a brain-like layered structure model using an RGD-gellan gum-based bioink (RGD-GG) encapsulating primary cortical neural cells. The authors reported a porous model structure (RGD-GG) that supported cell proliferation and network formation [48, 51, 118]. Gu and co-workers [86] reported on a novel neural mini-tissue construct using human cortical neuronal cells (NSCs) embedded in a polysaccharide hydrogel comprising carboxymethyl-chitosan, agarose and alginate. The system contained a homogeneous cell distribution that displayed good cell viability as well as in-situ differentiation of the NSCs [112, 118]. 3D Bioprinting technologies are appealing for neural tissue model fabrication and have been used to fabricate complex models for the study of 3D neurite outgrowth and elongation [119], human cortex design and engineering [120] as well as glioblastoma modelling [121].

*4.3.2.2 Microfluidics*

Microfluidic devices have been employed for *in vitro* studies in the development of multi tissue/interface structures to examine cortical brain structures providing a platform for drug development [122] with a focus on fabrication of microfluidic devices able to provide adequate solutions for the modelling of complex neural micro-environment such as the blood-brain-barrier (BBB) facilitating screening of BBB-targeting drugs in neurological diseases [48, 123]. Crucially, the parallel microgrooves commonly used in microfluidics to connect different tissue compartments offer ideal structures for neuron functionality and activity enabling enclosures for axon extension and pathfinding towards a target situated in adjacent chambers (**Fig. 3d**) [124]. Recent studies have focused in generating models that could facilitate communication between neural cells and other tissue, for the development of models for cancer metastasis and neuromuscular junction [125, 126]. As a result, information gained from studies on CNS axonal injury and regeneration on-a-chip [127], indicate microfluidics provides an unparalleled tool for human pain modelling, delivering unique *in vitro* platforms that enable the separation of axons from cell bodies and their localized treatment (e.g., drug testing and injury) through chamber compartmentalisation. In support of such an approach, Vysokov and colleagues reported a microfluidic system to model the pain synapse and investigate the role of voltage gated sodium channels (NaVs) in synaptic transmission [128]. Nevertheless, *in vitro* pain modelling remains poorly explored. It will undoubtedly fall on researchers to harness microfluidics in combination with other biotechnological tools such as optogenetic and multi electrode array (MEA) devices to advance neuro-pathophysiology research.

*4.3.2.3 Organoids*

Organoid models offer an attractive approach for the study of neural development and the testing of new treatments for neurological diseases [67]. The seminal work of Lancaster and co-workers provided a method to produce 3D cerebral human iPSCs-derived organoids [129] with the potential to apply organoids to model aspects of human neurodevelopment and neurological diseases. Subsequent work led to protocols to reproduce organoids derived from specific brain regions. Jo and co-workers [130], reported on the differentiation of human pluripotent stem cells into a large multicellular organoid-like structure, producing organoids with distinct layers of neuronal cells expressing genetic markers of human midbrain.
In the nervous system, pain stimuli trigger sensory neurons, which in turn transmit the signal to neurons in the dorsal horn of the spinal cord. An organoid model to recapitulate *in vitro* such an initial step of the nociceptive circuitry has been recently reported [131]. Ao and colleagues used an organoid-on-chip device made of a 3D printed holder with a porous polycarbonate membrane, in which a dorsal spinal cord organoid with both sensory neurons and dorsal spinal cord interneurons was maintained at the air-liquid interface. The ability to produce nociceptive responses was validated upon stimulation with known pain-evoking substances, including mustard oil, capsaicin, velvet ant venom as well as temperature increase to trigger the thermosensitive nociceptive pathway. The effect of pain relievers, including cannabinoids, could be observed in this model, confirming a platform for functional pain modelling *in vitro* [131].

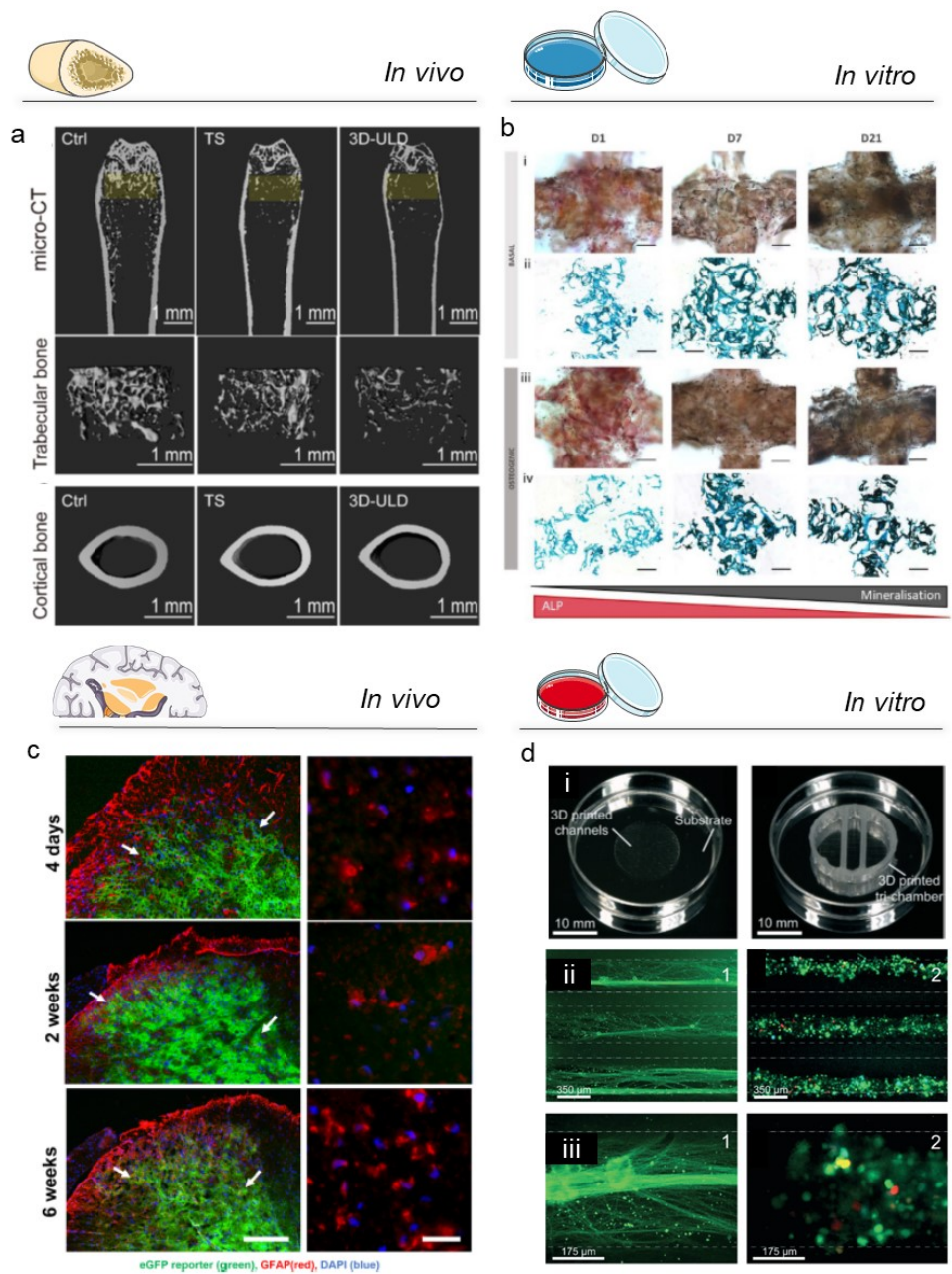

**Fig. 3** Bone and neuro models. (a) *In vivo*. A novel osteoporotic mouse model developed using an innovative movable and non-invasive unloading device (ULD). Micro-CT scan images (yellow highlighted analysis region) of the trabecular and cortical structures of mice femurs from control, tail suspension and 3D-ULD groups; adapted from [73] Copyright © 2021. This is an open-access article distributed under the terms of the Creative Commons Attribution License (CC BY). (b) *In vitro*. Harnessing nanoclay-based material, a 3D mineralising micro-environment for HBMSCs to proliferate and differentiate can studies *in vitro*, capable of developing a functional bone model in 21 days. ALP and Von Kossa staining for HBMSC-laden 3D bioprinted scaffolds cultured in basal and osteogenic conditioned media. Adapted from [89]; Copyright © 2020. This is an open-access article distributed under the terms of the Creative Commons Attribution License (CC BY). (c) *In vivo*. Model for the study of functional treatment for the crossing of blood-brain-barrier (BBB) to inhibit central reporter gene expression and study glial signalling to alleviate chronic pain. Oxytetracycline (Oxy) suppressed peripheral transgene expression. GFP-positive reported gene expression (green) observed extensively in the spinal cord tissue (left panels), while not localised in the sciatic nerve (right panels) at different time points. Glial fibrillary acidic protein (GFAP)-positive glial cells and eGFP expression was found to co-localise in specific regions – white arrows. Adapted from [105]. Copyright © 2022. This is an open-access article distributed under the terms of the Creative Commons Attribution License (CC BY). (d) *In vitro*. 3D printed nerve system on a chip. (i) 3D printed device comprising silicone microchannel for axonal guidance. (ii) Superior Cervical Ganglia (SCG) neurons with green-labelled tau protein aligned within the microchannel. Triple channels with self-assembled network of Schwann cells stained with PRV brainbow. (iii) Close-up images of above-mentioned detailed micrographs. Adapted from [124]. This article is licensed under a Creative Commons Attribution 3.0 Unported Licence

# 5. 3D models for mimicking bone pain

Bone pain models have preferentially involved the use of *in vivo* platforms for the monitoring of disease progression and therapeutic efficacy analysis for bone pain treatment and despite obvious limitations arising from species differences, offer a translational route to clinical application (**Table 1**). Osteoporotic, osteoarthritic and cancer-related bone fractures are examined in pre-clinical models for the study of bone pain with a focus on the generation of functional *in vivo* or *in vitro* models to recapitulate the mechanism underlying the disease conditions, to develop effective treatment strategies [132].

**Table 1.** Bone pain models with associated therapeutics findings or planned studies.

| Model | Animal | Therapy/Study | REF |
|---|---|---|---|
| *Cancer bone pain* | Rats with implantation of MAT B III (adenocarcinoma) cells | Blockade of IL-6 | [145] |
| | Mice with injection of Lewis lung cancer (LLC) | Adeno-associated virus shANXA3 (AAV-shANXA3) | [146] |
| | Mice subject to chemotherapy-induced peripheral neuropathy | Hsp90 inhibitors | [147] |
| | SD rats with Walker256 tumour tibial injection | TRPA1 antisense oligodeoxynucleotide delivery | [143] |
| | BALB/cAnNHsd mice with 66.1 breast cancer cells | Angiotensin-(1-7) (Ang-(1-7)) administration | [148] |
| | Mice with injection of Lewis lung cancer (LLC) | Spinal VEGF-A/VEGFR2 signalling blocked by intrathecal injection of the VEGF-A antibody or the specific VEGFR2 inhibitor ZM323881. | [149] |
| *Inflammatory-mediated bone pain* | Rats with carrageenan-induced inflammation | Artemin sequestration | [150] |
| *Osteoarthritic bone pain model* | Male Sprague-Dawley rats with unilateral intra-articular injection of monosodium iodoacetate | Evaluation of physical activity intensity and incidence on bone pain | [151] |
| *Bone afferent nerve* | Male Sprague-Dawley rats | Piezo2 knockdown with antisense oligodeoxynucleotides | [152] |
| | Mice with injection of dextran-biotin | Anterograde tracing study via injections of dextran–biotin | [153] |
| | Male Sprague-Dawley rats | Retrograde tracing and electrophysiological *in vivo* recording | [138] |

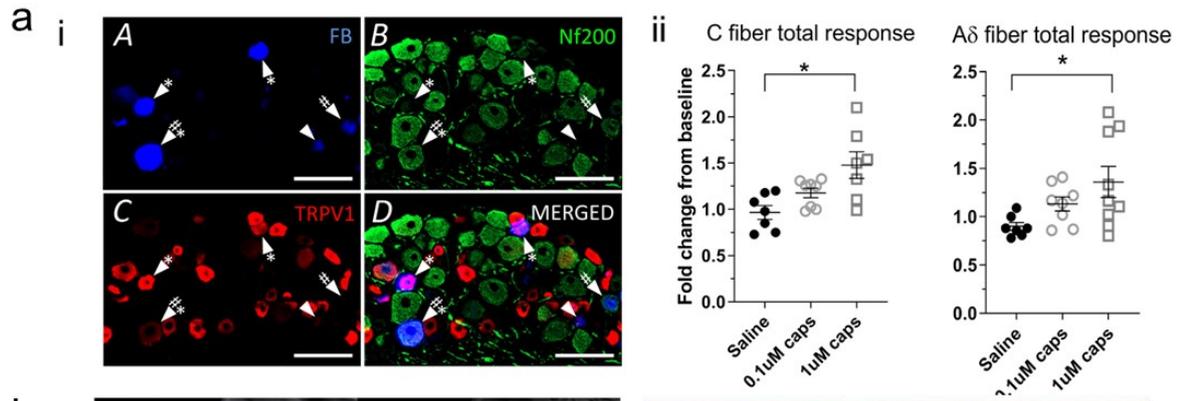
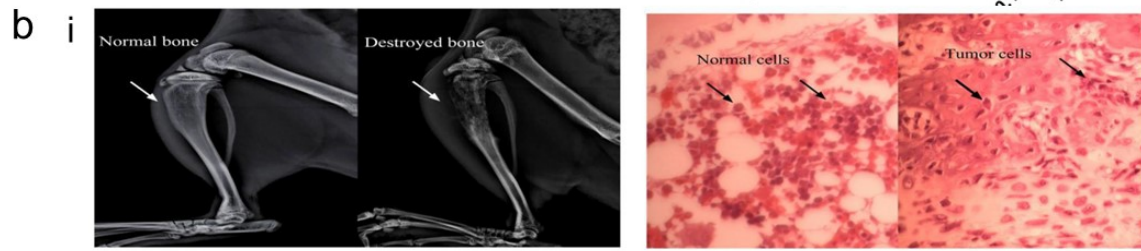
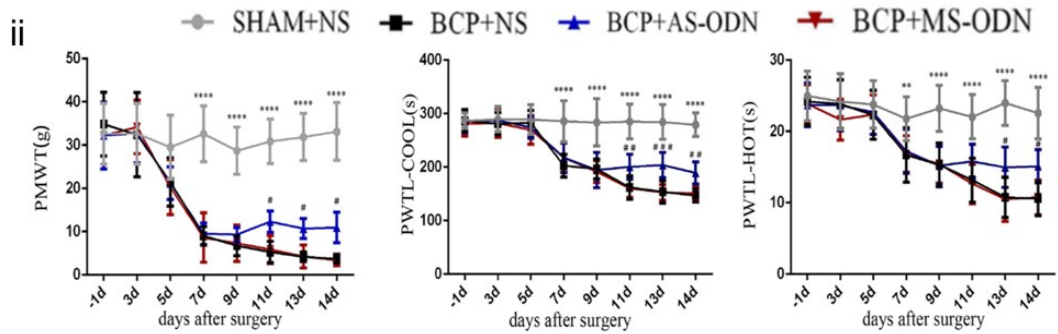
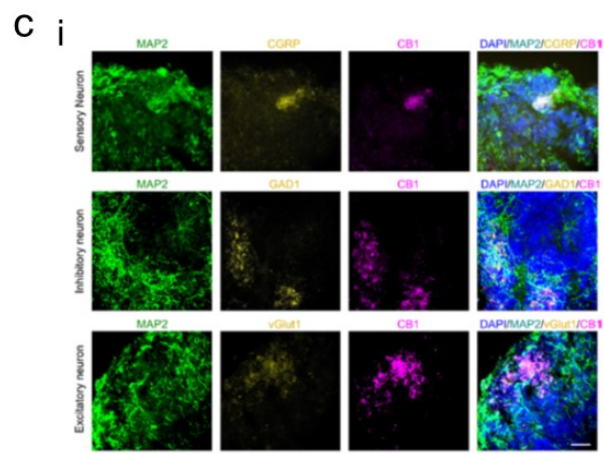
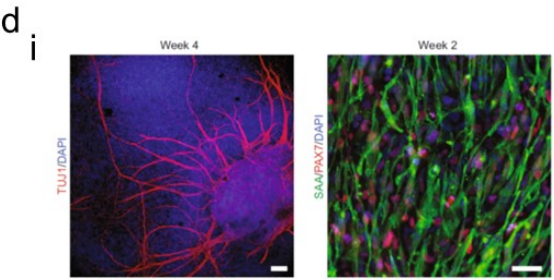
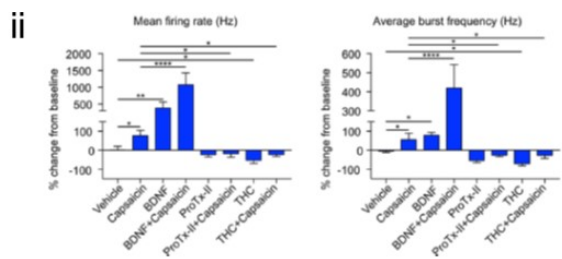
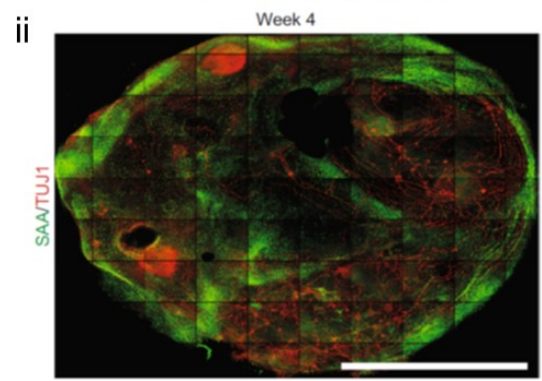

**Fig. 4** Bone pain models. (a) Bone pain *in vivo* model - TRPV1 expression. (i) L3 DRG section highlighting NF200+ neurons, TRPV1+ bone afferent neurons. (ii) The stimulation with 0.1 μM and 1 μM capsaicin increased the discharge frequency of both small (C fibre) and medium (Aδ fibre) amplitude. Adapted with permission from [138]. Copyright 2019. (b) Bone cancer pain model. Destruction of bone tissue (i) confirmed by X-ray imaging and histological analysis, confirming the presence of tumour cells within the marrow cavity. Paw mechanical withdrawal threshold (PMWT) and paw withdrawal thermal latency (PWTL) with cold and hot surfaces investigated in bone cancer rat model to evaluate the pain response *in vivo*. Relative expression of mRNA and Transient receptor potential ankyrin 1 (TRPA1) was enhanced in bone cancer pain animal model. Targeting via TRPA1 antisense oligodeoxynucleotide (AS-ODN) relieved PMWT and PWTL. Adapted from [143]. Copyright © 2021 under the terms of the Creative Commons Attribution License (CC BY). (c) Human spinal organoids in a chip (i) Positive staining for CB1 expression, for sensory, inhibitory, and excitatory neurons (CGRP+, GAD1+ and vGlut1+, respectively). ii) Response to capsaicin and electrical stimulation of spinal organoids towards nociceptive modulation, with enhanced mean firing rate and average burst frequency for BDNF and capsaicin stimulated group. Adapted with permission from [131]. Copyright 2022 American Chemical Society. (d) Human sensorimotor organoids based on (i) TUJ1+ neurons and sarcomeric α-actinin (SAA)+ myocytes can include both cell types (ii) and observed to be functional over 4 weeks Adapted from [144]. Creative Commons CC BY

## 5.1 Fracture models.

Bone fracture in rodent models is performed using tailored protocols where the selected bone segment is broken using a defined force [132–134]. The most common fracture model, developed by Bonnarens and Einhorn [135] provides a closed mid-shaft femoral fracture in a rat in which femur is stabilized with an intramedullary Steinman pin or Kirschner (K) wire and a diaphyseal fracture created using the Einhorn drop-weight apparatus [133, 135] .The model has proved highly reproducible, although is not suitable for the analysis of healing in delayed unions given the high healing rate typically observed in rodents. An alternative method reported by Tägil and co-workers [136] used a femoral osteotomy, with the periosteum removed around the fracture site, creating an open fracture model [133, 136]. This technique allows for a significant impairment of bone healing and favours the observation of the healing phenomenon.

Osteoarthritis and osteoporosis animal models have been engineered to fully recapitulate the skeletal discomfort and thus enable study of the underlying mechanisms and evaluation of treatment approaches for clinical translation. Osteoporotic models can be induced following ovariectomy in female mice, recreating a postmenopausal osteoporotic model with related bone fractures. However, the model is not representative of the human osteoporotic fracture as the mineral density of mice bone is lower than found in human bone [134]. Osteoarthritic pain induction, surgical interruption of the anterior cruciate ligament (ACL) and posterior cruciate ligament (PCL) in mouse have all been shown to be efficacious in the reproduction of osteoarthritic pain models. Such techniques allow the development of the pathology in an animal model with comparable developmental features observed in the human condition [132, 137].

To date, *in vivo* fracture models capable of replicating pain mechanisms have not been identified. Morgan and co-workers [138] reported the study of activation of Transient receptor potential cation channel subfamily V member 1 (TRPV1) expressed in bone afferent neurons (**Fig. 4a**) although could not recapitulate the complexity of bone fracture following the exclusion of crucial components such as the vascular network in damaged skeletal tissue.

## 5.2 Cancer bone pain models.

Cancer bone pain models have been extensively produced *in vivo*. The majority of the models involve the use of rodents in which bone metastases have been induced by localised cancerous cells injection. One of the first mouse models was proposed by Arguello and collaborators [139] in which intracardiac injections of myeloma cells were given to study the mechanism of bone localisation of blood-borne cancer cells. However, the pathological state was so severe the model precluded analysis of the pain experienced by the animal [132]. Schwei and colleagues [140] proposed an alternative model using a femur intramedullary injection of tumour cells to obtain a reproducible and localized tumour mass [28, 132]. The model was more effective in simulating the situation of cancer bone pain, recreating spontaneous pain and mechanical allodynia, enabling a comparison with the untreated limb [132]. Other studies have focused on *ex vivo* or *in vitro* reproduction of bone cancer using bone tissue explants or through biofabrication technologies. Nordstrand and colleagues [141] developed a co-culture system with physically separated murine calvarial explants and prostate cancer cells to study bone remodelling activity in the presence of cancer cells without being in direct contact with bone tissue [141, 142]. Established cancer bone pain models have found significant traction for the screening of novel drugs and therapies often targeting specific proteins and disrupting pathways involved in pain mechanisms (**Fig. 4b**). Approaches include the Transient receptor potential ankyrin 1 (TRPA1), recently targeted by Liu and co-workers [143] using TRPA1 antagonist and antisense oligodeoxynucleotide. Nevertheless, poor translation to human pathophysiology remains impairing pre-clinical drug screening and crucially, validation.

### 5.3 *In vitro* bone pain models

There remains an absence of *in vitro* bone pain models with only a few reports detailing either fracture healing approaches without the involvement of neural fibres, or the exclusive use of functional nociceptive models. A recent approach has harnessed spinal organoids, 3D printing and microfluidic technologies to demonstrate the stable generation of an *in vitro* pain model (**Fig. 4c**) [131]. However, the use of hESCs, the inability of include multiple cell types (e.g., BMSCs, myocytes) and the lack of further investigation are still limitations in translational potential of such a drug screen platform. In essence incorporation of multiple cell types, recapitulating a functional tissue interface, is a prerequisite to study bone pain mechanisms. This was illustrated by Pereira and colleagues [144] with the generation of a sensorimotor organoid model comprising both neurons (TUJ1+) and myocytes (SAA+) derived from iPSCs culture (**Fig. 4d**). The presence of sensory neurons and muscle cells offer a stable 3D platform approach, *in vitro,* for the investigation of pain. An obvious next step would be the inclusion of skeletal cell populations to provide additional perspective in the musculoskeletal pain paradigm.

## 6. Summary, Challenges and Future Perspectives

Bone pain discomfort arising as a consequence of trauma or skeletal diseases affect patient quality of life. Despite decades of research, the underlying mechanisms of pain affecting diseased or damaged bone tissue is to be fully elucidated. There remains an unmet need to generate functional models able to recapitulate the complex temporal cascade of biological events involved in bone pain development that could aid development of new therapeutic modalities for clinical pain management and drug screen research.

To date, animals have been routinely used as the perceived most reliable model for the study of bone pain mechanisms and therapy development. However, (i) ethical issues, (ii) species variability and (iii) intrinsic differences to human physiology limit routinely the translation of novel drug treatments. TE offers a third way to address some of these issues, providing specific tools and models that harness relevant cell types, scaffolds together with biofabrication/microfluidic and biotechnology strategies to create new approaches to model human bone pain. *In vitro* models offer significant potential and new vistas in research in simulating both pathological and bone pain conditions, helping to reduce, refine and, ultimately, replace animal models for the testing of bone pain treatments. Current research in bone pain modelling seeks to reproduce, in *vivo*, the pathological state and to subsequently induce pain with a platform comparable to the human bone lesion. A central and rather obvious limitation in the field remains the inability to recapitulate physiological bone innervation as a consequence of, often oversimplified models, lacking the essential requisite of different cell types necessary to recapitulate the bone-related pain cascade. Of course, these are not trivial issues, and the co-culture of bone and nerve cells is challenging. The possibility to engineer a micro-environment that can support chemical stimuli for cell survival, proliferation and signalling of pain, as well as mechanical stimuli due to external forces and matrix constraints, is hypothesised and yet, remains, by us and others in the field to be demonstrated. Therein lies the challenge and need for collaborative synthesis of methodologies across traditional boundaries of skeletal, nerve and indeed pain research.

The future is bright, interdisciplinary research harnessing life sciences, tissue engineers and clinicians and the raft of new methodologies for imaging to "omics" and biotechnology platforms auger well in the development of a third way, new avenues in the exploration of pain relief therapies, clearly needed for an aging demographic. Thus, new *in vitro,* models are anticipated, harnessing TE approaches that will lead to the prevention of bone pain with and widespread benefit across the global healthcare systems and, ultimately, patients.

## Acknowledgements

The authors have no conflicts of interest to declare. GC acknowledge funding from AIRC Aldi Fellowship under grant agreement No. 25412. ROCO acknowledges financial support from the Biotechnology and Biological Sciences Research Council (BB/P017711/1) and the UK Regenerative Medicine Platform "Acellular / Smart Materials – 3D Architecture" (MR/R015651/1). MR and AC acknowledge University of Pennsylvania Orphan Disease Center in partnership with the Fibrous Dysplasia Foundation (MDBR-21-110-FD) and Sapienza University RM120172B8BF5C15, RM118164289636F0). Figures were generated using Servier Medical Art (by Servier, licensed under a Creative Commons Attribution 3.0 Unported License)